\documentclass[prl,aps,epsf,showpacs,twocolumn]{revtex4-1}
\usepackage{times}
\usepackage{graphicx}
\usepackage{float}
\usepackage{latexsym,amsmath,amssymb,bm,euscript}

\usepackage{color}
\usepackage{bm}
\usepackage{units}
\usepackage{graphicx}
\usepackage{bbold}
\usepackage{tikz}
\usepackage{graphicx}% Include figure files
\usepackage{subfigure}
\usepackage{bm}% bold math
\usetikzlibrary{calc}
\usetikzlibrary{arrows}
\usepackage{hyperref}
\usepackage{comment}
 
\usepackage{color,soul}

\DeclareMathOperator{\im}{Im}

\usepackage{hyperref}
\makeatletter
\newcommand*{\rom}[1]{\expandafter\@slowromancap\romannumeral #1@}
\makeatother

\begin{document}

\title{Quantum critical nematic fluctuations and spin excitation anisotropy in iron pnictides}% Force line breaks with \\

\author{Chia-Chuan Liu$^{1}$}

\author{Elihu Abrahams$^{2}$}
\email{deceased}

\author{Qimiao Si$^{1}$}

\affiliation{
$^1$Department of Physics and Astronomy, Rice Center for Quantum Materials, Rice University, Houston, Texas 77005, USA\\
$^2$Department of Physics and Astronomy, University of California Los Angeles, Los Angeles, CA 90095, USA
} 

\begin{abstract}
Quantum criticality in iron pnictides involves both the nematic and antiferromagnetic degrees of freedom,
but the relationship between the two types of fluctuations has yet to be clarified.
Here we study this problem 
 in the presence of a small external uniaxial
potential, which breaks the $C_4$-symmetry in the B$_{1g}$ sector.
We establish 
a non-perturbative
identity that connects the spin excitation anisotropy, which is the difference of the dynamical spin 
susceptibilities at $\vec{Q}_1=\left(\pi,0\right)$ and $\vec{Q}_2=\left(0,\pi\right)$, with the dynamical magnetic susceptibility 
and static nematic susceptibility. 
Using
 this identity, we introduce a scaling procedure to determine the {\it dynamical} 
nematic susceptibility in the quantum critical regime, and illustrate the procedure for the case of the optimally Ni-doped 
BaFe$_2$As$_2$
[Y. Song \textit{et al.}, Phys. Rev. B \textbf{92},  180504 (2015)].
The implications of our results for the overall physics of the iron-based superconductors are discussed.
\end{abstract}

\maketitle

Iron-based superconductors have presented many intriguing and often puzzling properties \cite{Hosono,
Johnston_AP:2010,Wang_Sci:2011,Dai_RMP:2015,Si_NRM:2016,Hirschfeld_CRP:2016}. 
Among these is 
the onset of the tetragonal-to-orthorhombic structural phase transition at a temperature just above or at the antiferromagnetic (AF)
phase transition \cite{Pengcheng-nature}. 
When they are split, the region between the two transitions 
is called a nematic phase, where the $C_4$ tetragonal symmetry is broken 
while the $O\left(3\right)$ spin rotational symmetry is preserved. 
It has been well established that the nematic transition is driven by electron correlations, 
with B$_{1g}$ anisotropies in 
electronic, orbital and magnetic properties \cite{Chu-Science,Ming-PNAS,Lu-Science}.
Several channels are entwined in the nematic correlations,
including spin \cite{Si-PNAS,Fang-ising,Xu-ising,Willa2019}, electronic \cite{Littlewood,Weng-electron}
and
orbital\cite{Kruger-orbital,Weicheng-orbital} degrees of freedom.

\begin{figure} [b!]
\centering
\hspace{0.2 in}
\subfigure[ $\:\vec{Q}_1=\left(\pi,0\right)$]{
\includegraphics[scale=0.3]{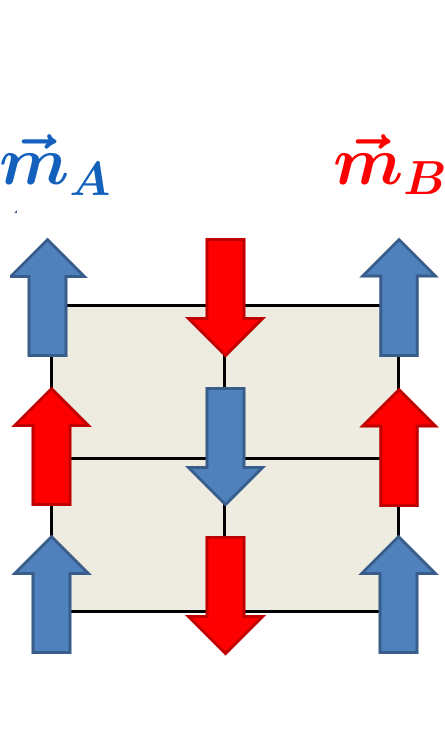}\label{fig:magpatterna}
}
\subfigure[ $\:\vec{Q}_2=\left(0,\pi\right)$]{\label{fig:magpatternb}
\includegraphics[scale=0.3]{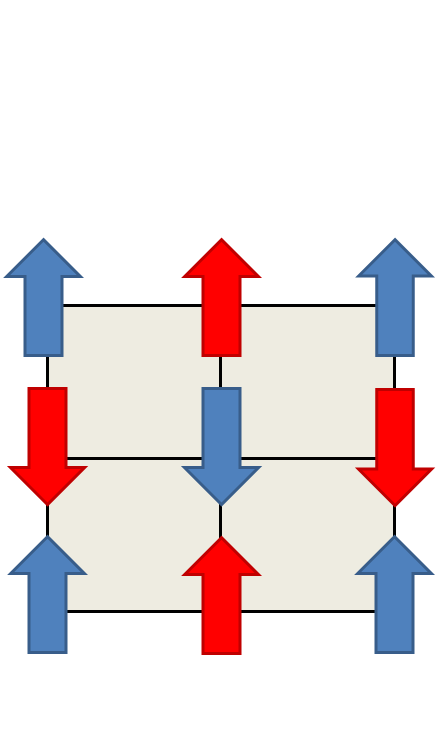}}
\subfigure[]{
\includegraphics[scale=0.25]{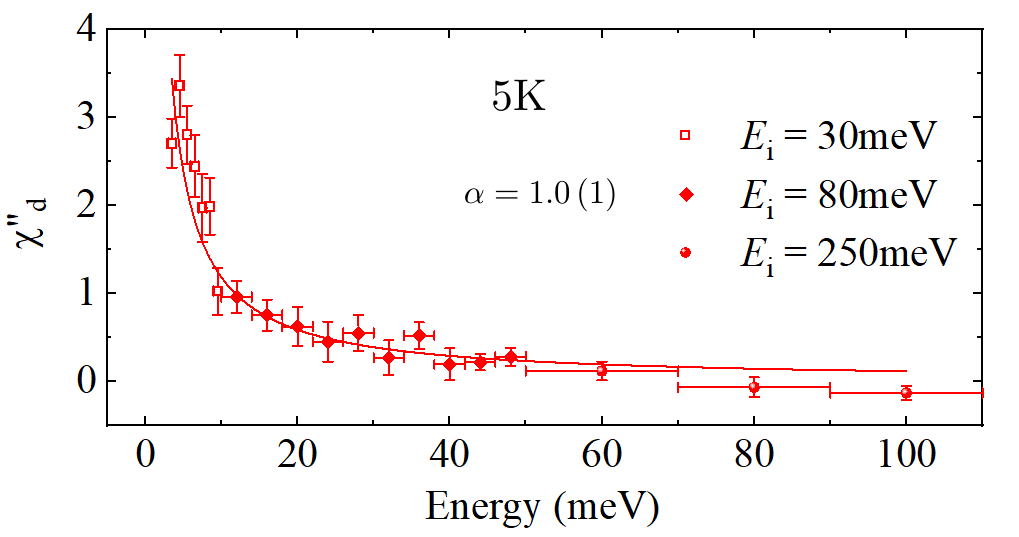}\label{fig:spindiff}}
\subfigure[]{
\includegraphics[scale=0.25]{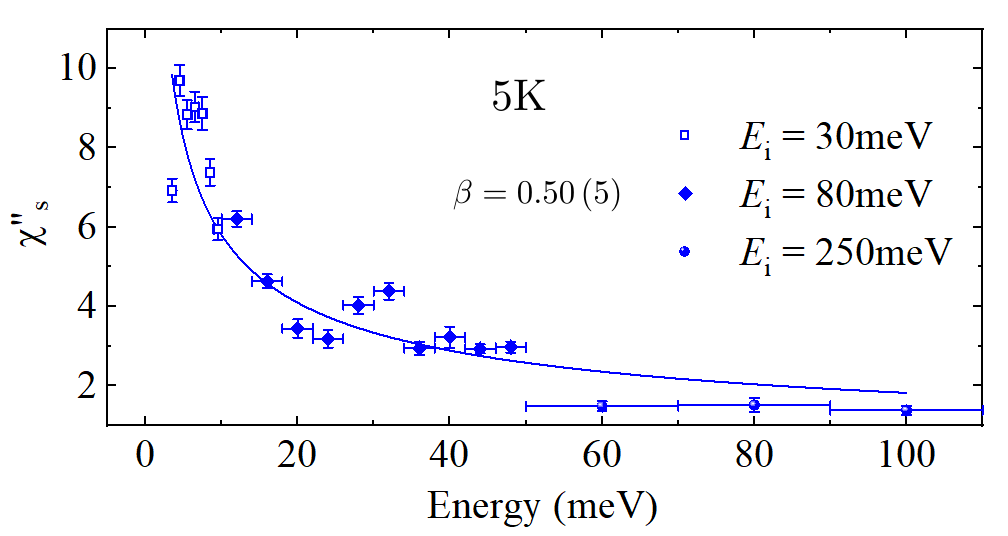}\label{fig:spinsum}}
\caption{\label{fig1}
(a) The spin configurations of the ground state of the parent iron-based superconductors with 
ordering wave vector $\vec{Q}_1=\left(\pi,0\right)$ or (b) $\vec{Q}_2=\left(0,\pi\right)$. 
The blue and red arrows denote the spins forming the staggered magnetizations on the sublattices A and B, respectively. 
Also shown are the energy dependences of 
(c)
the imaginary part of the spin excitation anisotropy $\chi''_{d}=\chi''\left(\vec{Q}_1\right)-\chi''\left(\vec{Q}_2\right)$
vs. energy
 and (d) the dynamical magnetic susceptibility $\chi''_{s}=\chi''\left(\vec{Q}_1\right)+\chi''\left(\vec{Q}_2\right)$ in 
 BaFe$_{2-x}$Ni$_{x}$As$_2$ measured by inelastic neutron scattering 
 experiments \cite{YuSpinex} near the optimal doping $x=x_c\approx 0.1$;
 the former (latter) is fit in the power law form ${E^{-\alpha}}$ (${E^{-\beta}}$) with the exponent $\alpha\cong 1.0$ 
 ($\beta\cong 0.5$).}
\end{figure}

One way to study the relationship between the nematic and other electronic channels
is to consider the quantum critical regime, where the critical
singularities 
can be isolated from regular contributions. 
The parent ground state of the iron-pnictide superconductors
is an AF state with the ordering wave vector $\vec{Q}_1=\left(\pi,0\right)$ or $\vec{Q}_2=\left(0,\pi\right)$. 
Their spatial patterns
are shown in Fig. \ref{fig:magpatterna} and \ref{fig:magpatternb}. 
The AF state breaks not only the usual $O\left(3\right)$ spin rotational symmetry, 
but also a $Z_2$ symmetry between the $\vec{Q}_1$  and $\vec{Q}_2$ magnetic state. 
In iron pnictides, the bad-metal behavior \cite{Si-PRL,Qazilbash-nature} motivated a theoretical proposal 
for the electronic excitations into coherent and incoherent parts. 
The tuning of the coherent electron weight was proposed to give rise to concurrent quantum criticality in
both 
the $\left(\pi,0\right)$ AF and Ising-nematic channels \cite{Si-PNAS,footnote}.
The existence of quantum criticality has been most extensively evidenced by experiments in
BaFe$_2$As$_2$
with P-for-As doping to the regime of optimal superconductivity
\cite{Jiang2009,Kasahara-prb,Nakai-prl,Analytis-nature,Hayes-nature}.

A defining characteristic of quantum criticality is 
the inherent mixing of statics and dynamics.
Singular magnetic responses in the quantum critical regime have been observed through
dynamical measurements
at both the optimally P-for-As- and Ni-for-Fe-doped
BaFe$_2$As$_2$ \cite{DHu2018,YuSpinex}. 
Singular nematic responses in the quantum critical regime have also been observed over 
a variety of optimally doped iron pnictides \cite{Kuo2016}, albeit in DC measurements.
The comparison already demonstrates the concurrent nature of the quantum criticality in the magnetic and 
nematic channels \cite{footnote2}.
However, to elucidate the relationship between the singular nematic and magnetic responses
and, by extension, for the purpose of analyzing the influence that these channels may have on the optimized
superconductivity,
it would be desirable to determine the (${\bf q}$-dependent) dynamical nematic susceptibility
in the quantum critical regime.
In general, such low-energy dynamical nematic susceptibility is not readily accessible experimentally. 

In this Letter, we address these issues by exploiting the relationship between the dynamical nematic susceptibility
and spin excitation anisotropy.
The latter, defined as 
the difference of the dynamical spin susceptibilities 
at $\vec{Q}_1=\left(\pi,0\right)$ and $\vec{Q}_2=\left(0,\pi\right)$,
under a uniaxial strain that breaks the $C_4$ symmetry in $B_{1g}$ channel,
has been measured
by inelastic neutron scattering experiments in the optimally doped iron pnictides
\cite{YuSpinex,Lu-Science,Luo-prl}.
We analyze the singular part of the dynamical responses in both the O(3) AF and Z$_2$ nematic 
sectors.
By performing a linear response calculation,
we establish a general
--and non-perturbative--
 identity [Eq.~(\ref{relation})] among the spin excitation anisotropy, the dynamical magnetic susceptibility, 
and the nematic susceptibility.
The
identity holds regardless of the microscopic mechanism of the nematicity.
Based on a scaling analysis, we further show how this identity can be 
used to explore the properties
of a quantum critical point (QCP),
 where both the magnetic and nematic channels are 
 concurrently critical. Through the scaling procedure, we
extract the dynamical nematic susceptibility from the spin excitation anisotropy,
and also determine the dynamic exponent $z$ and
the scaling dimension of the 
nematic order parameter $d_{\Delta}$.
The procedure is illustrated in the context of the inelastic
neutron scattering results for the optimally Ni-doped BaFe$_2$Si$_2$ 
under an external stress
\cite{YuSpinex}, which are summarized in 
Figs. \ref{fig:spindiff} and \ref{fig:spinsum}.

\textit{Spin excitation anisotropy and nematic susceptibility:~~}
The spin excitation anisotropy $\chi_{d}\left(\omega\right)$ and the dynamical magnetic susceptibility $\chi_{s}\left(\omega\right)$ are 
defined as the difference and summation of the dynamical spin susceptibility $\chi\left(\vec{q},\omega\right)$ between 
the two ordering wave vector $\vec{Q}_1=\left(\pi,0\right)$ and $\vec{Q}_2=\left(0,\pi\right)$, respectively:
\begin{eqnarray}
\chi_{s}\left(\omega\right) &\equiv&  \chi\left(\vec{Q}_1,\omega\right)+\chi\left(\vec{Q}_2,\omega\right)
\label{spinsumdef} \\
\chi_{d}\left(\omega\right) &\equiv & \chi\left(\vec{Q}_1,\omega\right)-\chi\left(\vec{Q}_2,\omega\right)
\label{spinexdef}
\end{eqnarray}

On symmetry grounds, the spin excitation anisotropy $\chi_d\left(\omega\right)$ 
should be related to the nematic fluctuations,
since it measures the degree of 
asymmetry of the magnetic fluctuations between
the two wave vectors $\vec{Q}_1$ and $\vec{Q}_2$.  
However, the precise relation 
has not been considered before. 

To proceed, we consider the problem in the presence of an external uniaxial stress, and focus on the effect of the induced strain in the $B_{1g}$ channel, which can couple to different kinds of microscopic degree of freedom such as electronic, orbital, or spin. 
Integrating out
the strain degree of freedom leads to a coupling between the stress and the nematic order parameter field  $S_{\lambda,\Delta}=\lambda\int^{\beta}_0 d\tau\int d^2x\Delta$, where $\lambda$ is a coupling constant that depends on the strength of the external stress and $\Delta$ is the nematic order parameter field. 
Depending on the mechanism for the nematicity, the nematic order parameter field $\Delta$ corresponds to different kinds of microscopic degree of freedom. For instance, in the spin-driven nematicity, it can be expressed in terms of the bilinear $\vec{m}_A\cdot\vec{m}_B$, where $\vec{m}_A$ and $\vec{m}_B$ are the Neel order parameter fields 
 on the sublattices A and B, respectively, as shown in Fig. \ref{fig:magpatterna}. 
 On the other hand, for the orbital-driven nematicity, it can be represented
 by
  $n_{xz}-n_{yz}$, where $n_{xz}$ and $n_{yz}$ are the occupations of the $d_{xz}$ and $d_{yz}$ Fe-orbitals, respectively.

We perform a linear response calculation by expanding only the small uniaxial term $S_{\lambda,\Delta}$ 
that results from the external uniaxial stress. A detailed analysis, given in SM, shows that:
\begin{equation}\label{relation0}
\begin{aligned}
&\chi_{s}\left(\omega\right)\equiv \chi\left(\vec{Q}_1 ,\omega\right)+\chi\left(\vec{Q}_2,\omega\right)=\overline{\chi}_{m}\left(0,\omega\right)+O\left(\lambda^2\right)
\end{aligned}
\end{equation}
and 
\begin{equation}\label{relation}
\begin{aligned}
&\chi_{d}\left(\omega\right)\equiv \chi\left(\vec{Q}_1 ,\omega\right)-\chi\left(\vec{Q}_2,\omega\right)\\
%&=\lambda\overline{V}\left(0,\omega\right)\overline{\chi}_{m_A}\left(0,\omega\right)\overline{\chi}_{\Delta}\left(0,0\right)\overline{\chi}_{m_B}\left(0,-\omega\right)+O\left(\lambda^2\right)\\
&=\lambda\overline{V}\left(0,\omega\right)\overline{\chi}^2_{m}\left(0,\omega\right)\overline{\chi}_{\Delta}\left(0,0\right)+O\left(\lambda^2\right)
\end{aligned}
\end{equation}
where 
$\overline{\chi}_{m}\left(q,\omega\right)\equiv \overline{\chi}_{m_A}\left(q,\omega\right)
=\overline{\chi}_{m_B}\left(q,\omega\right)$ is the magnetic propagator, $\overline{\chi}_{\Delta}\left(q,\omega\right)$ 
is the nematic propagator, and 
$\overline{V}$ is the irreducible vertex function involving two external magnetic order parameter fields $\vec{m}_A$ 
and $\vec{m}_B$ and one nematic order parameter field $\Delta$ \cite{footnote3}.
Here, an overline denotes the full correlation and vertex functions in the absence of the external stress. For both of the Eq.(\ref{relation0}) and Eq.(\ref{relation}), 
we assume the symmetry
$\vec{m}^2_A\leftrightarrow\vec{m}^2_B$, which is generally valid when there is no charge density order \cite{Rong2017}. 
The modified version of 
the identities when this symmetry is invalid can be found in the SM.

\begin{figure}[h!]
\centering
\includegraphics[scale=0.18]{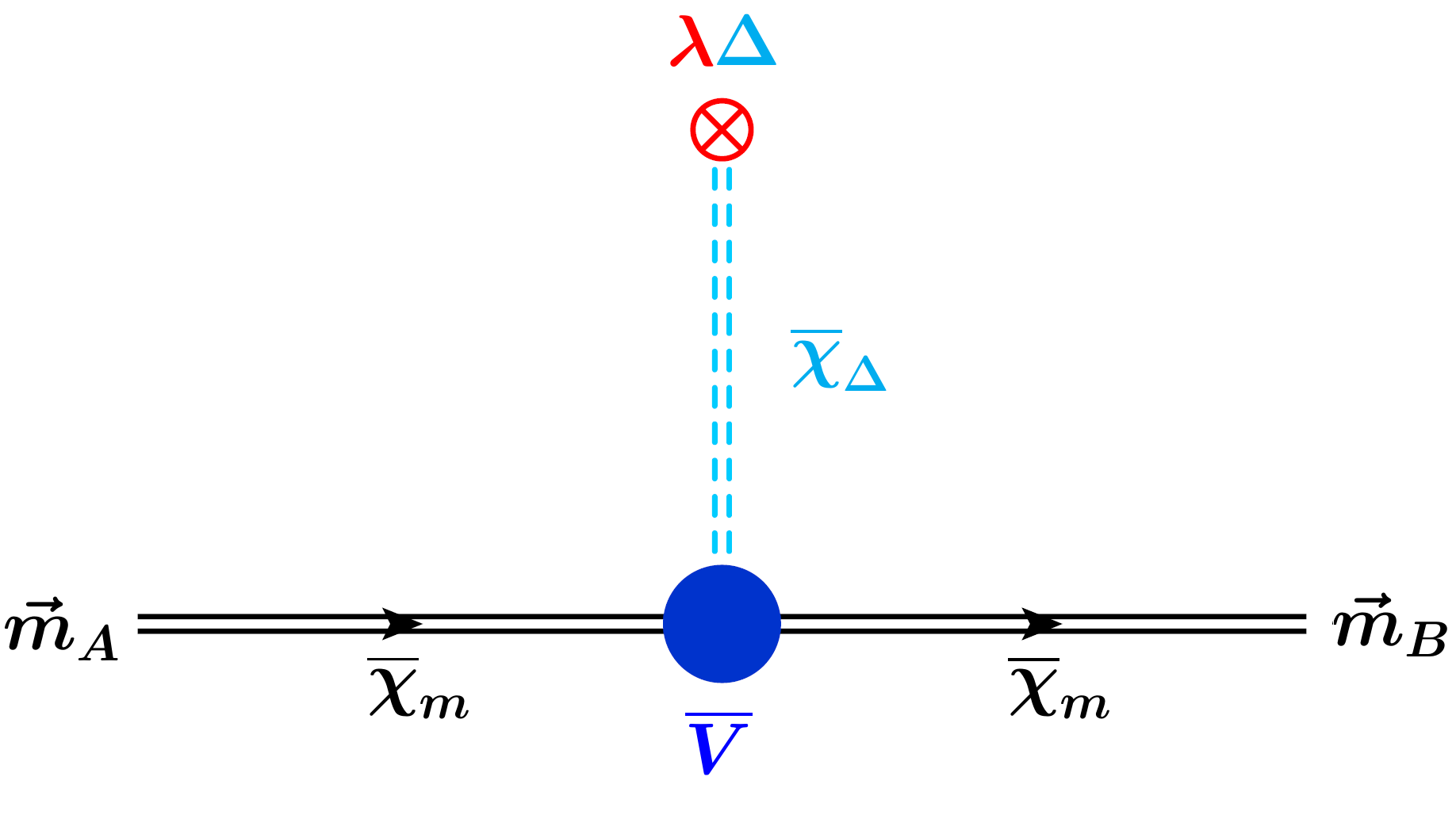}
\caption{The diagrammatic representation of the identity Eq.~(\ref{relation}). The double black line and double cyan dashed line denote
 the  
 magnetic propagator $\overline{\chi}_{m}$ and nematic propagator $\overline{\chi}_{\Delta}$, respectively. The blue circle is the  vertex function $\overline{V}$, and the red cross small circle is the external $C_4$ symmetry breaking potential.}
\label{spinexrela}
\end{figure}

The nematic order parameter field $\Delta$ has the same symmetry as
the magnetic bilinear $\vec{m}_A\cdot\vec{m}_B$; thus, they must be coupled to each other.
 Microscopically, this coupling is straightforward
 when $\Delta$ is constructed by the magnetic degrees of freedom;
 otherwise, it is determined by some higher-order processes, which is 
captured by the bare form of the vertex function $\overline{V}$. In other words, the identity Eq.~(\ref{relation}) reflects the symmetry property of the spin excitation anisotropy $\chi_{d}\left(\omega\right)$ (\ref{spinexdef}), and  holds regardless of the microscopic model. 
This identity will play a central role in the following analysis. The diagrammatic representation of this identity 
is shown in Fig. \ref{spinexrela}.

\textit{Scaling analysis:~~}
We now apply the identity, Eq.~(\ref{relation}), to extract the nematic susceptibility from the spin excitation anisotropy. 
Our focus is on the singular parts of these quantities in the quantum critical regime. At this point, we focus on the case that 
 both the AF and nematic channels are critical concurrently; the conditions under which this concurrent nature develops will be discussed below.

Due to the scale invariance in the quantum critical regime,
the irreducible two-point correlation function $\overline{\chi}_{m}\left(0,\omega\right)$
 and the irreducible vertex function $\overline{V}\left(0,\omega\right)$
  should have
  a power law form 
 with specific exponents  Therefore, we expect the spin excitation anisotropy 
 $\chi_d\left(\omega\right)$ to also 
show a power law 
 with a specific exponent.

To derive these exponents, we carry out a scaling analysis of the irreducible vertex functions, 
using the generating functional $\Gamma\left(m,\Delta\right)$ \cite{Peskinbook,Justin}:
\begin{equation}\begin{aligned}
&\Gamma\left(m,\Delta\right)=\sum_{n_m,n_{\Delta}}\frac{1}{n_m!n_{\Delta}!}\int_{\lbrace q\rbrace,\lbrace p\rbrace}\int_{\lbrace \omega\rbrace,\lbrace \nu\rbrace}\\
&\times\left( \prod_{i=1}^{n_m}m_i\prod_{j=1}^{n_\Delta}\Delta_j \right)\Gamma^{n_m,n_\Delta}\left(\lbrace q\rbrace,\lbrace p\rbrace,\lbrace\omega\rbrace,\lbrace\nu\rbrace \right)
\end{aligned}
\end{equation}
where we have defined the abbreviated notations 
$ \int_{\lbrace q\rbrace,\lbrace p\rbrace}\equiv \int \prod_{i=1}^{n_m} d^Dq_i \prod_{j=1}^{n_\Delta} d^Dp_j\delta\left(\sum_{i}^{n_m}q_i+\sum_{j}^{n_{\Delta}}p_j\right)$ and $ \int_{\lbrace \omega\rbrace,\lbrace \nu\rbrace}\equiv \int \prod_{i=1}^{n_m} d\omega_i \prod_{j=1}^{n_\Delta} d\nu_j\delta\left(\sum_{i}^{n_m}\omega_i+\sum_{j}^{n_{\Delta}}\nu_j\right)$, with
$D$ being the spatial dimensionality. In addition,
we have introduced $\Gamma^{n_m,n_{\Delta}}$ to represent
 an irreducible vertex function with $n_m$ and $n_{\Delta}$
external magnetic and nematic order parameter fields, respectively. 

In the quantum critical regime, 
under the 
rescaling
 $q\rightarrow  e^{-l}q$, $\omega\rightarrow e^{-zl}\omega$, 
the magnetic and
  nematic order parameter fields transform according to
 $m\rightarrow e^{-d_{m}l}m$,  $\Delta\rightarrow e^{-d_{\Delta}l}\Delta$,
  where $d_m$ and $d_{\Delta}$ are their respective scaling dimensions.
Since $\Gamma\left(m,\Delta\right)$ is a dimensionless quantity, 
   the irreducible vertex function $\Gamma^{n_m,n_{\Delta}}$ must satisfy \cite{Justin,Kleinert,Subir}:
\begin{equation}\label{dimenirre}
\begin{aligned}
&\Gamma^{n_{m},n_{\Delta}}\left(q,\omega\right)=\\
&e^{-
%\left[n_{m}\left( d_{m}+D+z\right)+n_{\Delta}\left(d_{\Delta}+D+z\right)-(D+z)\right]
d_{\Gamma}l}\Gamma^{n_{m},n_{\Delta}}\left(qe^{-l},\omega e^{-zl}\right)
\end{aligned}
\end{equation} 
where 
$d_{\Gamma} =
% \left[
n_{m}\left( d_{m}+D+z\right)+n_{\Delta}\left(d_{\Delta}+D+z\right)-(D+z)
%\right]
$,
and $z$ is the 
dynamic
exponent \cite{depend}.

The scaling dimension of the magnetic order parameter $d_m$ can be expressed in term of the dynamical exponent $z$ and the anomalous dimension $\eta$ as \cite{dimchoice}:
\begin{equation}\label{scaledimmag}
d_m=-\frac{D+z+2-\eta}{2} .
\end{equation}
Because $\eta$, the anomalous dimension,
is typically small, and for the purpose of the demonstration of our analysis, we will carry out our analysis assuming 
$\eta\cong 0$;
what happens when $\eta\neq 0$ is shown in the SM.
It then follows from Eq. (\ref{dimenirre})
that:
\begin{equation}\label{dimenirre2}
\begin{aligned}
&\Gamma^{n_{m},n_{\Delta}}\left(q,\omega\right)=\\
&e^{\left[n_{m}-n_{\Delta}\left( d_{\Delta}+D+z\right)-(D+z)\left(\frac{n_{m}}{2}-1\right)\right]l}\Gamma^{n_{m},n_{\Delta}}\left(qe^{-l},\omega e^{-zl}\right)
\end{aligned}
\end{equation} 

The magnetic propagator
is determined by a two-point irreducible vertex function:
\begin{equation}
\begin{aligned}
&\overline{\chi}^{-1}_m\left(q,\omega\right)\equiv \Gamma^{n_{m}=2,n_{\Delta}=0}\left(q,\omega\right)=e^{2\left(l\right)}\overline{\chi}^{-1}_m\left(qe^{-l},\omega e^{-zl}\right)\\
&=q^{2} \overline{\chi}^{-1}_m\left(1,\omega q^{-z}\right)
\end{aligned}
\end{equation}
where in the last step we choose $l$ such that $qe^{-l}=1$.
This, in turn, implies:
\begin{equation}\label{magfre}
\begin{aligned}
&\overline{\chi}_m\left(0,\omega\right)=\chi_s\left(\omega\right)\sim \omega^{-\frac{2}{z}}
\end{aligned}
\end{equation}

The function
$\overline{V}$ appearing in Eq.~(\ref{relation}) (and Fig.~\ref{spinexrela}) is a
three-point irreducible vertex function. The scaling procedure leads to:
\begin{equation}\label{verscal}
\begin{aligned}
&\overline{V}\left(q,\omega\right)\equiv \Gamma^{n_{m}=2,n_{\Delta}=1}\left(q,\omega\right)\\
&=e^{\left[2-d_{\Delta}-(D+z)\right]l}\overline{V}\left(qe^{-l},\omega e^{-zl}\right)\\
&=q^{\left(2-d_{\Delta}-(D+z)\right)}\overline{V}\left(1,\omega q^{-z}\right)
\end{aligned}
\end{equation} 
In turn, this gives rise to 
the following frequency dependence:
\begin{equation}\label{verfre}
\begin{aligned}
&\overline{V}\left(0,\omega\right) \sim \omega^{\frac{2-d_{\Delta}-(D+z)}{z}}
\end{aligned}
\end{equation} 

Collecting all these, 
we can now determine from Eq.~(\ref{relation})
 the scaling form for the spin excitation anisotropy:
\begin{equation}\label{spinexanexp}
\begin{aligned}
&\chi_d\left(\omega\right) \sim \omega^{\frac{-2-d_{\Delta}-(D+z)}{z}}
\end{aligned}
\end{equation} 

Conversely, 
by measuring the singular parts in the energy dependence of the spin excitation anisotropy $\chi_{d}\left(\omega\right)$ 
and dynamical magnetic susceptibility $\chi_s\left(\omega\right)$ in the quantum critical regime,
we can
determine the dynamical exponent $z$ and the scaling dimension of the nematic order parameter $d_{\Delta}$ 
through the  Eq.~(\ref{magfre}) and ~(\ref{spinexanexp}) for a given spatial dimensionality $D$. In turn, we can determine the singular {\it dynamical}
properties of 
 the nematic degree of freedom, which we now turn to.

\textit{Dynamical nematic susceptibility:~~}
The
analysis of the dynamical nematic susceptibility, $\overline{\chi}_{\Delta}\left(0,\omega\right)$,
 seems like an
  impossible task
   given that 
the identity Eq.~(\ref{relation})
involves only the static nematic susceptibility $\overline{\chi}_{\Delta}\left(0,0\right)$. 
The key point is that 
 the irreducible vertex function $\overline{V}\left(0,\omega\right)$ couples 
 the nematic and magnetic order parameter fields,
 and captures 
the critical singularity in the {\it dynamical} nematic correlations.

To make this point clear, we note that, according to the scaling analysis, 
the critical part of the dynamical nematic susceptibility $\overline{\chi}_{\Delta}\left(q,\omega\right)$
obeys the following form:
\begin{equation}\label{nemscal}
\begin{aligned}
&\overline{\chi}^{-1}_{\Delta}\left(q,\omega\right)\equiv \Gamma^{n_{m}=0,n_{\Delta}=2}\left(\vec{q},\omega\right)\\
&=e^{-\left[2d_{\Delta}+(D+z)\right]l}\overline{\chi}^{-1}_{\Delta}\left(qe^{-l},\omega e^{-zl}\right)\\
&=q^{-\left(2d_{\Delta}+(D+z)\right)}\overline{\chi}^{-1}_{\Delta}\left(1,\omega q^{-z}\right)
\end{aligned}
\end{equation} 
This, in turn, implies the following result for
the dynamical nematic susceptibility $\overline{\chi}_{\Delta}\left(0,\omega\right)$:
\begin{equation}\label{nemfre}
\begin{aligned}
&\overline{\chi}_{\Delta}\left(0,\omega\right)\sim \omega^{\frac{\left(2d_{\Delta}+(D+z)\right)}{z}}
\end{aligned}
\end{equation} 

\textit{The case of quantum criticality in BaFe$_2$As$_2$ with optimal Ni-doping:~~}
In the Ni-doped BaFe$_{2-x}$Ni$_{x}$As$_2$, the singular energy dependences of the spin excitation anisotropy
and magnetic susceptibility were observed near the optimal doping $x=x_c\approx 0.1$ by inelastic  neutron scattering
experiments \cite{YuSpinex} as shown in the Figs. \ref{fig:spindiff} and \ref{fig:spinsum} 
(where the fact that $T=5K<T_S$ only affects the results
at the lowest measured frequencies, given that the nematic order at this doping is already weak).
The experimental data suggest that the spin excitation anisotropy $\chi_{d}\left(\omega\right)$ and dynamical magnetic susceptibility $\chi_{s}\left(\omega\right)$ are best-fitted in power laws with different exponents $\alpha$ and $\beta$, respectively:
\begin{equation}\label{expdata}
\begin{aligned}
&\chi_d\left(\omega\right)\sim \omega^{-\alpha}\\
&\chi_{s}\left(\omega\right)=\overline{\chi}_m\left(0,\omega\right)\sim \omega^{-\beta}\\
\end{aligned} 
\end{equation}
with $\alpha\cong 1.0$ and $\beta\cong 0.50$.

We show how
 the identity Eq.~(\ref{relation}) provides
a natural way to explore the physics behind the relation (\ref{expdata}). 
To see this, consider a general form of the nematic propagator suitable for the quantum critical regime:
\begin{equation}
\overline{\chi}^{-1}_{\Delta}\left(q,\omega\right)=b_1q^n+b_2\frac{\vert\omega\vert}{q^a}
\end{equation}
For this propagator, we must have:
\begin{equation}\label{aspropdy}
\begin{aligned}
&z=n+a\\
\end{aligned}
\end{equation}
and 
\begin{equation}\label{aspropnem}
\begin{aligned}
&d_{\Delta}=-\frac{D+z+n}{2}=-\frac{D+2n+a}{2}\\
\end{aligned}
\end{equation}

Substitute Eqs.~(\ref{aspropdy}) and (\ref{aspropnem}) into (\ref{magfre}) and (\ref{spinexanexp}), and then compare with the experimental results (\ref{expdata}), we can find that the quantum-critical nematic susceptibility is:
\begin{equation}\label{nempropfit}
\overline{\chi}^{-1}_{\Delta}\left(q,\omega\right)=b_1q^2+b_2\frac{\vert\omega\vert}{q^2}
\end{equation}
when the spatial dimensionality $D=2$.

\textit{Discussion:~~}

We now turn to discussing several important points.
First, 
we have stressed that the identity, Eq.\,(\ref{relation}), is a generic consequence of the symmetry property 
of the spin excitation anisotropy $\chi_d$, defined in Eq.\,(\ref{spinexdef}). 
Different microscopic models affect the precise form of the nematic order parameter and the vertex function, but
do not affect the relationship between
 the dynamical magnetic susceptibility, static nematic susceptibility
and the vertex function.
Depending on the origin of the 
nematicity \cite{Si-PNAS,Fang-ising,Xu-ising,Willa2019,Littlewood,Weng-electron,Kruger-orbital,Weicheng-orbital,Ferne-nature},
 the nematic order parameter field $\Delta$ here comprises different building blocks.

Second, our analysis has assumed that the quantum criticality is concurrent between the 
spin and nematic channels. This has been strongly supported by experimental measurements  
 over a substantial dynamical range in both the static nematic susceptibility and dynamical spin susceptibility. In BaFe$_2$(As$_{1-x}$P$_x$)$_2$, both channels of quantum criticality 
 \cite{DHu2018,YuSpinex,Kuo2016}
 appear at the P-doping of about $x=0.3$,
 which corresponds to optimized superconductivity.
This concurrent quantum criticality
was anticipated theoretically based on the spin-driven nematicity.
The latter corresponds to the effective Ginzburg-Landau action $S_0$, as specified in Eq.\,(S.21) of the Supplementary Material.
Analyses using a renormalization-group method \cite{Si-PNAS} and in a large-$N$ limit \cite{JWu-phase}
showed that this action exhibits the quantum criticality concurrent in the AF and nematic channels.

Third,
 the nematic propagator 
 (\ref{nempropfit})
contains 
a non-trivial critical dynamical term $\vert\omega\vert/q^2$.
To see the implication of this result,
 we consider the 
symmetry-dictated coupling between the  nematic
and antiferromagnetic channels,
$\Delta\vec{m}_A\vec{m}_B$,
within the effective action for the
spin-driven nematicity [see $S_0$ of Eq.\,(S.21) in SM].
While the presentation of a systematic large-$N$-based approach to study this coupling \cite{CCLiu-arxiv} 
is beyond the scope of the present work, we note that
the decay of the nematic field
into a pair of critical magnetic fluctuations  leads to an extra damping dynamical term:
\begin{equation}\label{imnemprop}
\im{\chi^{-1}_{\Delta}\left(\vec{q},i\omega\rightarrow \omega+i0^+\right)}\propto \frac{\omega}{q^2} \,
\end{equation}
which 
accounts for 
Eq.\,(\ref{nempropfit}) . 
This understanding, together with the underlying concurrency of the quantum criticality just discussed,
supports the spin-driven mechanism of the nematicity in the iron pnictides.

Finally, 
with the primary goal of our work being to determine the nematic propagator of the
 iron pnictides, we consider the successful application of our approach to the case of the optimally Ni-doped Ba122 iron pnictides 
 as a major advance. This is the only iron-based materials class in which the spin excitation anisotropy has been measured 
 over a sizable frequency range. Our work should motivate further such dynamical measurements in other iron pnictides, 
 such as the optimally P-doped 122 iron pnictides.
Still, given that the static nematic susceptibility has shown universal critical behavior across the iron pnictides
and chalcogenides \cite{Kuo2016},
it is likely that the dynamical nematic susceptibility we have determined applies across the iron-based superconductors.

\textit{Conclusion:~~}
To summarize, we have advanced a rigorous identity that shows how the singular component of the 
spin excitation anisotropy connects with its counterparts in both the nematic and dynamical 
magnetic susceptibilities.
This identity holds regardless of the microscopic mechanism for the nematicity and has allowed us to extract the critical properties from the experiments in an optimally doped iron pnictides under a 
 uniaxial strain, including several critical exponents and a singular 
 nematic susceptibility as a function of both frequency and wavevector.
Our approach allows us to
determine the dynamical nematic susceptibility, which is difficult to directly measure experimentally. The singular fluctuations in
both the nematic and magnetic channels appear in the regime of
optimized superconductivity within the iron-pnictide 
phase diagram. Thus, both are expected to influence the development of the superconductivity.

\acknowledgements 
The bulk of this work was carried out prior to the passing of one of the authors (E.A.), 
with whom the other two authors (C.-C.L. and Q.S.) are grateful for having had the opportunity to collaborate. 
We thank Ian Fisher, Ding Hu, Steven Kivelson, Jianda Wu, Rong Yu, and particularly Pengcheng Dai 
and Yu Song for useful discussions.
This work has been supported in part by the U.S. Department of Energy, Office of Science, Basic Energy
Sciences, under Award No. DE-SC0018197 and the Robert A. Welch Foundation Grant No. C-1411 (C.-C. L and Q.S.). 
One of us (Q.S.) acknowledges the hospitality of the Aspen Center for Physics, 
which is supported by the NSF (Grant No.  PHY-1607611).

\clearpage
%%%%%%%%%%%%%%%%%%%%%%%%%%%
\widetext

\setcounter{figure}{0}
\setcounter{equation}{0}
\makeatletter
\renewcommand{\thefigure}{S\@arabic\c@figure}
\renewcommand{\theequation}{S\arabic{equation}}
\renewcommand{\bibnumfmt}[1]{[S#1]}
\renewcommand{\citenumfont}[1]{S#1}

\begin{center}
\begin{large}
{\bf Supplemental Material}
\end{large}
\end{center}

%\begin{comment}

\renewcommand{\theequation}{S.\arabic{equation}}

\section{nematicity and the Spin excitation anisotropy } \label{sm:sea}

%\begin{comment}
The spin excitation anisotropy $\chi_{d}\left(\omega\right)$ and the dynamical magnetic susceptibility $\chi_{s}\left(\omega\right)$ is defined as:
\begin{equation}\label{sm:spinsumdefsm}
\chi_{s}\left(\omega\right)\equiv \chi\left(\vec{Q}_1,\omega\right)+\chi\left(\vec{Q}_2,\omega\right)
\end{equation}
\begin{equation}\label{spinexdefsm}
\chi_{d}\left(\omega\right)\equiv \chi\left(\vec{Q}_1,\omega\right)-\chi\left(\vec{Q}_2,\omega\right)
\end{equation}
where $\vec{Q}_1=\left(\pi,0\right)$ and $\vec{Q}_2=\left(0,\pi\right)$ are ordering wave vectors, and the dynamical spin susceptibility $\chi\left(\vec{q},\omega\right)$ is:
\begin{small}
\begin{equation}\label{spinsussm}
\begin{aligned}
&\chi\left(\vec{q},\omega\right)=\int d^2r'\int d^2r\int  d\tau e^{-i\vec{q}\cdot\left(\vec{r}'-\vec{r}\right)-i\omega\tau} \langle T_{\tau} \vec{S}\left(\vec{r},\tau\right)\cdot \vec{S}^*\left(\vec{r}',0\right)\rangle_S=\int  d\tau e^{-i\omega\tau} \langle T_{\tau} \vec{S}\left(\vec{q},\tau\right)\cdot \vec{S}^*\left(-\vec{q},0\right)\rangle_S\\
\end{aligned}
\end{equation}
\end{small}
and $\vec{S}\left(\vec{r},\tau\right)$ is the local spin operator.

	In order to probe the spin excitation anisotropy $\chi_{d}\left(\omega\right)$, a small external uniaxial stress is necessary to detwin the sample. Therefore, the expectation $\langle T_{\tau} \vec{S}\left(\vec{r},\tau\right)\cdot \vec{S}^*\left(\vec{r}',0\right)\rangle_S$ in Eq.(\ref{spinsussm}) is calculated under the action $S=S_0+S_{\lambda,\Delta}$.
Here,  $S_0$ is the intrinsic action in the absence of the external uniaxial stress;
its explicit form does not matter to our analysis,
although it must respect
some general symmetries as we will discuss below.

Our next step is to express the dynamical magnetic susceptibility $\chi_{s}\left(\omega\right)$ 
and the spin excitation anisotropy $\chi_{d}\left(\omega\right)$ 
in terms
 of the magnetic order parameter fields and the nematic order parameter field $\Delta$. From the usual definition of the $A$ and $B$ sublattices [{\it cf.} Fig.\,1(a,b) of the main text] \cite{Abrahams-Si_jpcm},
 it follows that the spin operator field $\vec{S}\left(\vec{q},\tau\right)$ at ordering wave vectors  $\vec{q}=\vec{Q}_{1,2}$ 
 can be expressed as $\vec{S}\left(\vec{Q}_{1},\tau\right)=\left[\vec{m}_A\left(\vec{q}=0,\tau\right)-\vec{m}_B\left(\vec{q}=0,\tau\right)\right]/2$ and  $\vec{S}\left(\vec{Q}_{2},\tau\right)=\left[\vec{m}_A\left(\vec{q}=0,\tau\right)+\vec{m}_B\left(\vec{q}=0,\tau\right)\right]/2$.
Here, $\vec{m}_A$ and $\vec{m}_B$
are  the coarse-grained Neel order parameter fields on each sublattices A and B respectively, 
and their spatial fluctuations are
small on the scale of the lattice constant.
 We can then 
express the spin excitation anisotropy $\chi_d\left(\omega\right)$ and dynamical magnetic 
susceptibility $\chi_s\left(\omega\right)$ in terms of the magnetic order parameter fields $\vec{m}_A$ and $\vec{m}_B$:
\begin{equation}\label{spinexmathsm}
\begin{aligned}
&\chi_{d}\left(\omega\right)\equiv \chi\left(\vec{Q}_1,\omega\right)-\chi\left(\vec{Q}_2,\omega\right)\\
&=-\frac{1}{2}\int d\tau e^{-i\omega\tau}\langle T_{\tau}\vec{m}_A\left(\vec{q}=0,\tau\right)\cdot\vec{m}_B\left(\vec{q}'=0,\tau'=0\right)\rangle_S-\frac{1}{2}\int d\tau e^{-i\omega\tau}\langle T_{\tau}\vec{m}_B\left(\vec{q}=0,\tau\right)\cdot\vec{m}_A\left(\vec{q}'=0,\tau'=0\right)\rangle_S\\
&=-\int d\tau e^{-i\omega\tau}\langle T_{\tau}\vec{m}_A\left(\vec{q}=0,\tau\right)\cdot\vec{m}_B\left(\vec{q}'=0,\tau'=0\right)\rangle_S\\
\end{aligned}
\end{equation}
and
\begin{equation}\label{spinsummathsm}
\begin{aligned}
&\chi_{s}\left(\omega\right)\equiv \chi\left(\vec{Q}_1,\omega\right)+\chi\left(\vec{Q}_2,\omega\right)\\
&=\frac{1}{2}\int d\tau e^{-i\omega\tau}\langle T_{\tau} \vec{m}_{A}\left(\vec{q}=0,\tau\right)\cdot\vec{m}_{A}\left(\vec{q}'=0,\tau'=0\right)\rangle_S +\frac{1}{2}\int d\tau e^{-i\omega\tau}\langle T_{\tau} \vec{m}_{B}\left(\vec{q}=0,\tau\right)\cdot\vec{m}_{B}\left(\vec{q}'=0,\tau'=0\right)\rangle_S\\
&=\int d\tau e^{-i\omega\tau}\langle T_{\tau} \vec{m}_{A/B}\left(\vec{q}=0,\tau\right)\cdot\vec{m}_{A/B}\left(\vec{q}'=0,\tau'=0\right)\rangle_S
\end{aligned}
\end{equation}
Here, in the last equality for 
Eq.\,(\ref{spinexmathsm}) and 
that for
 Eq.\,(\ref{spinsummathsm}), we
 have assumed that
 the intrinsic action $S_0$ respects the symmetry $\vec{m}^2_A\leftrightarrow \vec{m}^2_B$, and thus there is no condensation of the
 the field
  $\langle \vec{m}_A^2-\vec{m}_B^2\rangle $. 
  Since $\vec{m}_A^2-\vec{m}_B^2$ has the same symmetry as a charge density order, they
 must
 have a bilinear coupling; for
a system without a charge density order,
$\langle \vec{m}_A^2-\vec{m}_B^2\rangle $ vanishes.

The expectation values in Eqs.~(\ref{spinexmathsm}) and (\ref{spinsummathsm}) are calculated under the action $S=S_0+S_{\lambda,\Delta}$. 
We treat the uniaxial strain $S_{\lambda,\Delta}$ as a perturbation and expand the action with respect to $S_0$:

\begin{small}

\begin{equation}\label{expandsm}
\begin{aligned}
&\langle T_{\tau} \vec{m}_A\left(0,\tau\right)\cdot\vec{m}_A\left(0,0\right)\rangle_{S_0+S_{\lambda,\Delta}}= \langle T_{\tau} \vec{m}_A\left(0,\tau\right) \cdot\vec{m}_A\left(0,0\right)\rangle_{S_0}-\lambda \int d\tau' \langle T_{\tau} \vec{m}_A\left(0,\tau\right)\cdot \vec{m}_A\left(0,0\right) \Delta\left(0,\tau'\right)\rangle_{S_0}+O\left(\lambda^2\right)\\
&= \langle T_{\tau} \vec{m}_A\left(0,\tau\right) \cdot\vec{m}_A\left(0,0\right)\rangle_{S_0}+O\left(\lambda^2\right)=\int d\omega' e^{i\omega'\tau} \overline{\chi}_{m_A}\left(0,\omega'\right)+O\left(\lambda^2\right) \, .
\end{aligned}
\end{equation}
\end{small}
We have used $\langle T_{\tau} \vec{m}_A\left(0,\tau\right) \vec{m}_A\left(0,0\right) \Delta\left(0,\tau'\right)\rangle_{S_0}=0$, and the fact
that 
 $\overline{\chi}_{m_A}\left(\vec{q},\omega'\right)$ is the magnetic propagator in the momentum space. 
 Following Eq.~(\ref{spinsummathsm}) and Eq.~(\ref{expandsm}), we have:
\begin{equation}\label{relationsumsm}
\chi_s\left(\omega\right)=\frac{1}{2}\left( \overline{\chi}_{m_{A}}\left(0,\omega\right)+ \overline{\chi}_{m_{B}}\left(0,\omega\right)\right)= \overline{\chi}_{m}\left(0,\omega\right) \, .
\end{equation}
Here, in the last equality, we have used
   $\overline{\chi}_{m_{A}}\left(0,\omega\right)=\overline{\chi}_{m_{B}}\left(0,\omega\right)$ 
given that the symmetry $\vec{m}^2_A\leftrightarrow \vec{m}^2_B$ 
is respected in the absence of a charge density order,
and have defined $\overline{\chi}_{m}\left(0,\omega\right)\equiv \overline{\chi}_{m_{A}}\left(0,\omega\right)=\overline{\chi}_{m_{B}}\left(0,\omega\right)$.

We see that $\chi_{s}\left(\omega\right)$ is just the dynamical magnetic propagator $\overline{\chi}_{m}\left(0,\omega\right)$.
Similarly, for the spin excitation anisotropy $\chi_{d}\left(\omega\right)$:
\begin{small}

\begin{equation}\label{threepointsm}
\begin{aligned}
&\langle T_{\tau}\vec{m}_A\left(0,\tau\right)\cdot\vec{m}_B\left(0,0\right)\rangle_{S_0+S_{\lambda,\Delta}}= \langle T_{\tau} \vec{m}_A\left(0,\tau\right) \cdot\vec{m}_B\left(0,0\right)\rangle_{S_0}-\lambda\int d\tau' \langle \vec{m}_A\left(0,\tau\right)\cdot \vec{m}_B\left(0,0\right) \Delta\left(0,\tau'\right)\rangle_{S_0}+O\left(\lambda^2\right) \, ,
\end{aligned}
\end{equation}
\end{small}
where $\langle T_{\tau} \vec{m}_A\left(0,\tau\right)\cdot \vec{m}_B\left(0,0\right)\rangle_{S_0}\propto \langle \Delta_I\rangle_{S_0}\delta_{\tau,0} $,
with $\Delta_I$ being
the Ising-nematic order parameter field.
Now,
$\Delta_I$ must linearly couple with 
the nematic order parameter field $\Delta$, since they have the same symmetry. As a result, $\langle \Delta_I\rangle_{S_0}\propto \langle \Delta\rangle_{S_0}$, and thus $\langle T_{\tau} \vec{m}_A\left(0,\tau\right)\cdot \vec{m}_B\left(0,0\right)\rangle_{S_0}\propto \langle \Delta\rangle_{S_0}\delta_{\tau,0} $, which vanishes
at temperatures $T>T_S$, the nematic phase transition temperature, or
is negligible at
$T<T_S$ but near the nematic (classical or quantum)
critical point at which $ \langle \Delta\rangle_{S_0}\rightarrow 0$.
Our analysis will focus on these regimes, in which
 Eq.\,(\ref{threepointsm}) becomes:
\begin{small}

\begin{equation}\label{threepointsm2}
\begin{aligned}
&\langle T_{\tau}\vec{m}_A\left(0,\tau\right)\cdot\vec{m}_B\left(0,0\right)\rangle_{S_0+S_{\lambda,\Delta}}=
%C \langle \Delta\rangle_{S_0}\delta_{\tau,0}
-\lambda\int d\tau' \langle \vec{m}_A\left(0,\tau\right)\cdot \vec{m}_B\left(0,0\right) \Delta\left(0,\tau'\right)\rangle_{S_0}+O\left(\lambda^2\right)\\
\end{aligned}
\end{equation}
\end{small}

Because
any three-point correlation function can always be factorized as the product of suitable irreducible two-point correlation function and irreducible vertex function\cite{Peskinbook,Justin,Kleinert}, we have:
\begin{equation}\label{threepointfactorsm}
\begin{aligned}
& \int d\tau' \langle  T_{\tau}\vec{m}_A\left(0,\tau\right)\cdot\vec{m}_B\left(0,0\right) \Delta\left(0,\tau'\right)\rangle_{S_0}=\int d\omega' e^{i\omega'\tau} \overline{V}\left(0,\omega'\right)\overline{\chi}_{m_A}\left(0,\omega'\right)\overline{\chi}_{\Delta}\left(0,0\right)\overline{\chi}_{m_B}\left(0,-\omega'\right)
\end{aligned}
\end{equation}

As a result, we conclude:
\begin{equation}\label{relationsm}
\begin{aligned}
&\chi_{d}\left(\omega\right)\equiv \chi\left(\vec{Q}_1 ,\omega\right)-\chi\left(\vec{Q}_2,\omega\right)=
%-C\langle \Delta\rangle_{S_0}+
\lambda\overline{V}\left(0,\omega\right)\overline{\chi}_{m_A}\left(0,\omega\right)\overline{\chi}_{\Delta}\left(0,0\right)\overline{\chi}_{m_B}\left(0,-\omega\right)+O\left(\lambda^2\right)\\
&=\lambda\overline{V}\left(0,\omega\right)\overline{\chi}^2_{m}\left(0,\omega\right)\overline{\chi}_{\Delta}\left(0,0\right)+O\left(\lambda^2\right)
%-C\langle \Delta\rangle_{S_0}
\end{aligned}
\end{equation}
where 
 $\overline{\chi}_{\Delta}$ is the nematic propagator, and $\overline{V}$ is the  vertex function involving two external magnetic order parameter fields $\vec{m}_A$ and $\vec{m}_B$, and one nematic order parameter field $\Delta$. Again, we 
have used
 $\overline{\chi}_{m_A}\left(0,\omega\right)=\overline{\chi}_{m_B}\left(0,\omega\right)=\overline{\chi}_{m}\left(0,\omega\right)$.
In addition, it follows from
 the time reversal symmetry that
 $\overline{\chi}_{m}\left(0,\omega\right)=\overline{\chi}_{m}\left(0,-\omega\right)$. 
 We have then derived
  the identity (4)
 of  the main text.

Note that when we derive the identities (\ref{relationsumsm}) and (\ref{relationsm}), 
only the uniaxial strain term $S_{\lambda, \Delta}$ 
is treated perturbatively;
no other approximation has been made.
In other words, the identities (3) and (4)  in the main text
  are valid non-perturbatively
 as far as the intrinsic action
 $S_0$ is concerned. 

\section{Scaling analysis of spin excitation anisotropy when $\eta\neq 0$}\label{sm:anomalous}

In this section, we carry through the scaling analysis of the spin excitation anisotropy $\chi_d\left(\omega\right)$ 
and dynamical magnetic susceptibility $\chi_s\left(\omega\right)$ with a non-zero anomalous dimension $\eta$ for 
the magnetic order parameter field $\vec{m}$. The scaling dimension of the magnetic order parameter field is:
\begin{equation}
d_m=-\frac{D+z+2-\eta}{2}
\end{equation}
with $\eta\neq 0$.

Following a procedure 
in parallel to
what was presented in the main text, 
we find the frequency dependence of the magnetic propagator
 in the quantum critical regime to be:
\begin{equation}
\begin{aligned}
& \overline{\chi}_m\left(0,\omega\right)=\chi_s\left(\omega\right)\sim \omega^{-\frac{2-\eta}{z}}
\end{aligned}
\end{equation}
and the corresponding
 frequency dependence of the vertex function to be:
\begin{equation}\label{vertano}
\begin{aligned}
&\overline{V}\left(0,\omega\right) \sim \omega^{\frac{\left(2-\eta\right)-d_{\Delta}-(D+z)}{z}}
\end{aligned}
\end{equation}

Consequently, by the identity Eq.~(4) in the main text, the spin excitation anisotropy is:
\begin{equation}
\begin{aligned}
&\chi_d\left(\omega\right) \sim \omega^{\frac{-\left(2-\eta\right)-d_{\Delta}-(D+z)}{z}}
\end{aligned}
\end{equation} 
Therefore,  again we can 
determine
the dynamical exponent $z$ and the scaling dimension of the nematic order parameter $d_{\Delta}$ 
from
 the dynamical nematic susceptibility $\chi_{s}\left(\omega\right)$ and the spin excitation anisotropy $\chi_{d}\left(\omega\right)$ according to:
\begin{equation}
\begin{aligned}
&\frac{-\left(2-\eta\right)}{z} =\frac{\partial\ln{\chi_s\left(\omega\right)}}{\partial\ln{\omega}}\\
&\frac{-d_{\Delta}-(D+z)-\left(2-\eta\right)}{z} =\frac{\partial\ln{\chi_d\left(\omega\right)}}{\partial\ln{\omega}}\\
\end{aligned}
\end{equation} 

We are now in position to see the implications of
 the singular energy dependences of the spin excitation anisotropy and the dynamical magnetic susceptibility 
observed in the optimally Ni-doped BaFe$_2$As$_2$. Consider
the nematic propagator 
in a general form
suitable for
the quantum critical regime:
\begin{equation}
\overline{\chi}^{-1}_{\Delta}\left(q,\omega\right)=b_1q^n+b_2\frac{\vert\omega\vert}{q^a}
\end{equation}
by which we then know:
\begin{equation}
\begin{aligned}
&z=n+a\\
\end{aligned}
\end{equation}
and
\begin{equation}
\begin{aligned}
&d_{\Delta}=-\frac{D+z+n}{2}=-\frac{D+2n+a}{2}\\
\end{aligned}
\end{equation}

Comparing
 it with Eq.~(17) in the main text, it is straightforward to show that:
\begin{equation}
\begin{aligned}
&a=4-D-2\eta=2-2\eta\\
&n=2+2\eta
\end{aligned}
 \end{equation}

\section{Spin-driven nematicity}\label{sm:spin-driven}
 
The effective Ginzburg-Landau action for the spin-driven nematicity is as follows:
\begin{equation}\label{j1j2ginz}
\begin{aligned}
&S_0=S_2+S_4\\
&S_2=\sum_{q=\vec{q},i\omega_n}\Big\{G_0^{-1}\left(q\right)\left(\vert \vec{m}_{A}\left(q\right)\vert^2+\vert \vec{m}_{B}\left(q\right)\vert^2\right)+v\left(q^2_x-q^2_y\right)\vec{m}_A\left(q\right)\cdot\vec{m}_B\left(-q\right)\Big\}\\
%&+2v\left(q^2_x-q^2_y\right)\vec{m}_A\left(q\right)\cdot\vec{m}_B\left(-q\right)\Big\}\\
&S_4=\int^{\beta}_0 d\tau\int d^2 x\Big\{u_1\left(\vec{m}^2_A+\vec{m}_B^2\right)^2-u_I\left(\vec{m}_A\cdot\vec{m}_B\right)^2-u_2\left(\vec{m}^2_A-\vec{m}_B^2\right)^2\Big\} \, .
\end{aligned}
\end{equation}
where $\vec{m}_A$ and $\vec{m}_B$ are the Neel order parameter fields on the sublattices A and B, respectively, and $G_0^{-1}\left(\vec{q},i\omega_n\right)=r+\omega_n^2+c\vec{q}^2+\gamma\vert \omega_n\vert$ with 
the mass term $r$ and  
Landau damping term $\gamma\vert\omega_n\vert$ resulting from
the coherent
electronic excitations \cite{Si-PNAS}.

The implications of the Ginzburg-Landau action (\ref{j1j2ginz}) in the present context are discussed in the main text.


\begin{thebibliography}{}
\bibitem{footnote3}
Strictly speaking, the identity (\ref{relation}) holds at temperatures $T>T_S$, where $T_S$ is the nematic phase transition temperature, 
or $T<T_S$ but near the nematic (classical or quantum) critical point.
See SM for further details.
\bibitem{depend} 
Here, the scaling analysis is carried out
in the quantum critical regime.
\bibitem{dimchoice} Note that in Eqs.~(\ref{dimenirre}) and (\ref{scaledimmag}),
 we choose to count the dimension from the momentum space action.
\end{thebibliography}
\end{document}